\documentclass[twocolumn,showpacs,superscriptaddress,english,aps,prb]{revtex4}
\usepackage[T1]{fontenc}
\usepackage[latin9]{inputenc}
\usepackage{amstext}
\usepackage{graphicx}
\usepackage{amssymb}
\usepackage{bm}
\usepackage{babel}

\makeatletter

\providecommand{\LyX}{L\kern-.1667em\lower.25em\hbox{Y}\kern-.125emX\@}
\newcommand{\lyxmathsym}[1]{\ifmmode\begingroup\def\b@ld{bold}
  \text{\ifx\math@version\b@ld\bfseries\fi#1}\endgroup\else#1\fi}

\@ifundefined{textcolor}{}
{%
 \definecolor{BLACK}{gray}{0}
 \definecolor{WHITE}{gray}{1}
 \definecolor{RED}{rgb}{1,0,0}
 \definecolor{GREEN}{rgb}{0,1,0}
 \definecolor{BLUE}{rgb}{0,0,1}
 \definecolor{CYAN}{cmyk}{1,0,0,0}
 \definecolor{MAGENTA}{cmyk}{0,1,0,0}
 \definecolor{YELLOW}{cmyk}{0,0,1,0}
 }

\makeatother

\newcommand{\bQ}{\mbox{\boldmath$Q$}}

\begin{document}

\title{Continuous magnetic and structural phase transitions in Fe$_{1+y}$Te.}

\author{I.~A.~Zaliznyak}
\email{zaliznyak@bnl.gov}
\affiliation{CMPMSD, 
 Brookhaven National Laboratory, Upton, NY 11973 USA}
\author{Z.~J.~Xu}
\affiliation{CMPMSD, 
Brookhaven National Laboratory, Upton, NY 11973 USA}
\author{J.~S.~Wen}
\affiliation{CMPMSD, 
Brookhaven National Laboratory, Upton, NY 11973 USA}
\affiliation{Physics Department, University of California, Berkeley, CA 94720, USA}
\author{J.~M.~Tranquada}
\affiliation{CMPMSD, 
 Brookhaven National Laboratory, Upton, NY 11973 USA}
\author{G.~D.~Gu}
\affiliation{CMPMSD, 
Brookhaven National Laboratory, Upton, NY 11973 USA}
\author{V.~Solovyov}
\affiliation{CMPMSD, 
Brookhaven National Laboratory, Upton, NY 11973 USA}
\author{V.~N.~Glazkov}
\affiliation{P. Kapitza Institute for Physical Problems, Moscow, Russia}
\affiliation{Neutron Scattering and Magnetism Group, Laboratorium f\"{u}r Festk\"{o}rperphysik,
ETH H\"{o}nggerberg, Z\"{u}rich, Switzerland}
\author{A.~I.~Zheludev}
\affiliation{Neutron Scattering and Magnetism Group, Laboratorium f\"{u}r Festk\"{o}rperphysik,
ETH H\"{o}nggerberg, Z\"{u}rich, Switzerland}
\author{V.~O.~Garlea}
\affiliation{Oak Ridge National Laboratory, 1, Bethel Valley Road, Oak Ridge, Tennessee 37831, USA}
\author{M.~B.~Stone}
\affiliation{Oak Ridge National Laboratory, 1, Bethel Valley Road, Oak Ridge, Tennessee 37831, USA}

\begin{abstract}

We report a sequence of continuous phase transformations in iron telluride, Fe$_{1+y}$Te ($y \approx 0.1$), which is observed by combining neutron diffraction, magnetic susceptibility, and specific heat measurements on single crystal samples. While a gradual increase of magnetic scattering near the wave vector $(0.5,0,0.5)$ is seen below $T \approx 70$ K, a temperature where the discontinuous first order magneto-structural phase transition is found in systems with small $y$ ($\lesssim 0.06$), the reduction of the lattice symmetry in Fe$_{1.1}$Te only occurs at $T_s \approx 63$ K. Below $T_N \approx 57.5$ K the long-range magnetic order develops, whose incommensurate wave vector $\bQ_m$ varies with temperature. Finally, at $T_m \lesssim 45$ K the system enters the low-$T$ phase, where $\bQ_m$ is locked at $\approx (0.48,0,0.5)$.  We conclude that these instabilities are weak compared to the strength of the underlying interactions, and we suggest that the impact of the Fe interstitials on the transitions can be treated with random-field models.

\end{abstract}


\maketitle


\section{Introduction}

Iron telluride is an end member of the simplest chalcogenide family of iron-based high-temperature superconductors (HTSC), Fe$_{1+y}$Te$_{1-x}$Se$_{x}$. It becomes superconducting upon partial (or full) isoelectronic substitution of Te by Se.\cite{Yeh2008,Hsu2009,Wen2011} Although the highest critical temperature for FeTe$_{1-x}$Se$_x$ is only $T_c \approx 14.5$ K, it increases to above 30 K in K$_x$Fe$_2$Se$_2$, or under pressure.\cite{Margadonna2009,Guo2010}  The FeTe crystal structure consists of a continuous stacking of square-lattice layers of iron atoms, each sandwiched between the two half-density layers of bonding chalcogen atoms, which is the basic structural motif for all iron-based superconductors. The Te atoms, which tetrahedrally coordinate the Fe sites, occupy alternate checkerboard positions above and below the Fe layer, so that the resulting unit cell contains two formula units. In this quasi-two-dimensional structure, FeTe layers are held together only by weak Van der Waals forces. Crystallographic stability is improved if some amount of extra Fe atoms is incorporated between the layers, which has important consequences for the low-temperature phases observed in the Fe$_{1+y}$Te series, $0.02 < y < 0.18$.\cite{Bao2009,Rodriguez2011,Stock2011,Li2009,Martinelli2010,Liu2011}

Similarly to the cuprate and the ferro-pnictide HTSC families, the end member Fe$_{1+y}$Te has a magnetically ordered ground state and undergoes a structural distortion which lowers the high-temperature tetragonal (HTT) lattice symmetry.\cite{Kastner1998,LynnDai2009,LumsdenChristianson2010} The physics behind these low-temperature phases and their relation to the superconductivity are of great interest and have been the subjects of intense study.\cite{Nakajima2011,ZhaoDai2008,LiDai2009,ChuFisher2009,Rotundu2010,Nandi2010,Ni2010,Kim2011,Marty2011} Two general trends of the phase diagrams were established: (i) unless there is a first order magneto-structural transition, the lattice distortion usually occurs at a higher temperature ($T_s$) than the magnetic ordering ($T_N$), $T_s \gtrsim T_N$, and (ii) both $T_s$ and $T_N$ are reduced upon chemical substitution, so that both orders tend to disappear as the superconducting state develops. While this observation suggests a close connection between the magnetic ordering, the lattice distortion (LD), and the superconductivity, other studies indicate that there might be no causal relationship between these phenomena.\cite{Canfield2010,Katayama2010,Shimojima2010} In that case, they are simply different manifestations of the same complex physics underlying the electronic behavior of HTSC materials. The mechanisms by which composition affects the nature of these magnetic and structural transitions (first or second order, which occurs first), as well as their possible connections to superconductivity, are still not well understood.

Fe$_{1+y}$Te provides an opportunity for investigating this issue. It is non-superconducting, but recent neutron studies have discovered that both magnetism and the low-$T$ crystal structure are extremely sensitive to the Fe stoichiometry.\cite{Bao2009,Rodriguez2011} We must note that the crucial issue of controlling the Fe stoichiometry, $y$, is also a tedious one. In particular, we measured the iron content, $y$, in several representative small crystals using the inductively-coupled plasma (ICP) method, and then performed supplementary neutron powder diffraction (NPD) measurements on specimens obtained by grinding single-crystal pieces. \cite{PhaseDiagram2011} We found that $y$ obtained by refining the occupancy of the interstitial site in NPD is typically about 3\% higher than the chemical Fe content obtained with ICP, $y \equiv y_{ICP} \approx y_{NPD} - 0.03$, for $y \gtrsim 0.04$. Similar discrepancy was also observed by other groups. \cite{Stock} It might imply that a certain amount of Fe vacancies exist, along with the interstitials. Here, we use the chemical Fe content as measured by the ICP as the appropriate notation for $y$ in the Fe$_{1+y}$Te formula. This has to be kept in mind when comparing our results with other studies, such as in Refs \onlinecite{Bao2009,Rodriguez2011,Stock2011}, where different determinations of $y$ have been used. We keep their original notations when discussing these results. Finally, we have also found that different crystals from the same growth can have different $y$, thus calling for extreme caution when preparing powder specimens, as well as for the need of properly characterizing $y$ for each sample.

At low $y$ ($\lesssim 0.06$), Fe$_{1+y}$Te undergoes a first-order magneto-structural phase transition, where the HTT $P4/nmm$ lattice symmetry is reduced to the monoclinic $P21/m$, and a ``bi-collinear'' magnetic order appears with propagation vector $(1/2,0,1/2)$ in $P4/nmm$ reciprocal lattice units.\cite{Bao2009,Rodriguez2011,Stock2011,Li2009,Martinelli2010} The main peculiarity of the ``bi-collinear'' magnetic ordering in Fe$_{1+y}$Te materials compared to the simple bipartite antiferromagnetism of other HTSC families is that it does not agree with the Fermi surface nesting of itinerant electrons, which corresponds to the $(1/2,1/2)$ position in the $ab$-plane.\cite{Hanaguri2010,Xia2009,ZhangFeng2010,Kanigel2011} Hence, band structure calculations, which tend to account for the Fermi surface,\cite{Subedi2008,Zhang2009,Ma2009,HanSavrasov2009} encounter difficulties when confronted with the broad range of experimental observations \cite{Bao2009,Rodriguez2011,Stock2011,Li2009,Martinelli2010,Kanigel2011,Chen2009,HuPetrovic2009,Khasanov2009,Bendele2010,Babkevich2010,Lipscombe2011,Zaliznyak2011}. On the other hand, the ground-state ordered moment, $\langle\mu \rangle \lesssim 2\mu_B$, \cite{Bao2009,Rodriguez2011,Li2009,Martinelli2010} although larger than in parent ferropnictides, is significantly less than $4\mu_B$ ($\mu_B=$ Bohr magneton) expected in the ionic local-spin picture for Fe$^{2+}$ ($S=2$) in the Hund's state.\cite{HauleKotliar2009} While some theories postulate that Fe could be in the non-Hund, intermediate-spin S=1 state, \cite{Turner2009} such an assumption falls way short of accounting for the paramagnetic moment $\mu_{eff} \approx 4\mu_B$ implicated in the Curie-Weiss behavior above 100 K \cite{Chen2009,HuPetrovic2009}. Thus, in Fe$_{1+y}$Te we can explore the interaction of weak magnetic order with lattice and orbital degrees of freedom in a representative structure of Fe-based superconductors, disentangled from the effects of Fermi surface nesting and superconductivity.

The phase diagram for $y \gtrsim 0.06$ is still controversial. Bao {\it et al.}\cite{Bao2009} who first discovered the orthorhombic $Pnmm$ phase with an incommensurate magnetic order for $y \approx 0.14$, initially suggested that the incommemnsurability varies linearly with $y$, and the transition is first order. A more recent NPD study \cite{Rodriguez2011} suggested that the first-order transition to a $P21/m$, ``bi-collinear'' commensurate phase survives until $y \approx 0.12$, at which point a mixed phase is observed. At higher $y$, the low-$T$ phase is orthorhombic, with coexisting long-range helimagnetism and short-range spin-density-wave (SDW) order.

Here, we report studies of well-characterized single crystals of Fe$_{1+y}$Te, $y = 0.10(1)$, with a variety of techniques, which establish a sequence of continuous phase transitions. It starts with a structural distortion at $T_s = 63(1)$K, which is followed by slightly incommensurate magnetic order at $T_N = 57.5(5)$ K. This implies a multicritical point, $y_c$ on the $(y,T)$ phase diagram of Fe$_{1+y}$Te, with $0.06 \lesssim y_c \lesssim 0.1$, where the first-order magneto-structural phase transition turns into a sequence of continuous ones. While according to Ref.  \onlinecite{Rodriguez2011} our samples belong to the grey ``mixed'' phase, we find no evidence for the mixed character. Instead, we find a well-defined sequence of phases as a function of temperature. This leads us to favor the idea that samples studied in Ref. \onlinecite{Rodriguez2011} were mixtures of different stoichiometries, $y$, among other possibilities suggested by the authors of that study. This conclusion is further supported by our more recent additional NPD and bulk measurements of the $(y,T)$ phase diagram of Fe$_{1+y}$Te. \cite{PhaseDiagram2011} They suggest that multicritical point at $y_c$ is indeed an intrinsic property of this phase diagram. Hence, our findings provide a direct connection between the composition phase diagram of Fe$_{1+y}$Te and that of  BaFe$_2$As$_2$ (122) derived materials, where a similar multicritical point has recently been focus of considerable work.\cite{ChuFisher2009,Rotundu2010,Nandi2010,Ni2010,Kim2011,Marty2011}

\section{Experimental procedure}

The Fe$_{1.1}$Te crystals were grown by the horizontal Bridgman method.\cite{Wen2011}  The crystal used for the neutron-scattering study had a mass $m = 18.45$~g and a mosaic of $2.2^\circ$ full-width at half-maximum (FWHM).

Our neutron measurements were carried out using the ARCS direct-geometry, time-of-flight (TOF) spectrometer at the Spallation Neutron Source, Oak Ridge National Laboratory. The instrument was operated in the Laue-diffraction-type mode,\cite{Bozin2009} where a quasi-white neutron beam with a broad band of incident neutron energies centered around $E_i \approx 300$ meV was selected by the pre-monochromating $T_0$ chopper. The Fe$_{1.1}$Te crystal was mounted on an aluminum holder attached to the cold head of the closed-cycle refrigerator in the ARCS evacuated scattering chamber. The crystal's $c$-axis was aligned in the horizontal plane, at $\approx 45^\circ$ to the incident neutron beam, while the $a$-axis was at about $24^\circ$ to the horizontal plane. The detector signal is dominated by the elastic processes (diffraction), where the scattering angle is determined by the incident neutron energy (or wavelength, $\lambda_i$) and the $d$-spacing of the set of crystal planes involved in reflection, in accordance with Bragg's law, $\lambda_i = 2d \sin \theta$. Such a measurement is particularly well suited for studying the relative temperature evolution of structural and magnetic scattering, which are both present in the diffraction pattern at each $T$.

Bulk magnetization, $M$, and heat capacity, $C_p$, were measured on several single crystals with masses from 7.5~mg to 26 ~mg, cleaved from the same growth boule as the large crystal used for neutron studies. $C_p$ was measured using the relaxation method implemented in the Physical Properties Measurement System (PPMS) by Quantum Design (QD). Crystals were attached to a silver sample holder using apiezon grease, whose contribution to $C_p$ was measured and subtracted. Its uncertainty was the main source of the dominant systematic error. The non-magnetic lattice phonon specific heat was estimated using two different algorithms. First, we used the $C_p(T)$ of non-magnetic ZnTe,\cite{Gavrichev2002} with temperature re-scaled by the ratio of the effective Debye temperatures, $\theta_D$, of Fe$_{1.1}$Te and ZnTe, obtained by fitting their $C_p(T)$ in the $T \gtrsim \theta_D$ range to a single Debye function (short-dashed lines in Fig. \ref{Fig2:Cp}). Secondly, we used an equal-weight sum of two Debye and one Einstein functions, as described in Ref. \onlinecite{Tsurkan_EPJ2011} for the FeSe$_{0.5}$Te$_{0.5}$ case, re-scaled in a similar way (solid lines in Fig. \ref{Fig2:Cp}).

The static magnetic susceptibility, $\chi = M/H$, was obtained from $M$ measured using a QD Magnetic Properties Measurement System, in a dc magnetic field $\mu_0 H = 0.1$~T applied in the $ab$-plane, $\chi_{ab}$, or along the $c$-axis, $\chi_{c}$. The iron content, $y = 0.10(1)$, was measured in these and several other representative small crystals, using the inductively-coupled plasma method. The $10\%$ error bar on $y$ results from the scatter of different ICP measurements. We performed supplementary neutron powder diffraction measurements on specimens obtained by grinding single-crystal pieces \cite{PhaseDiagram2011} and established that $y$ obtained by refining the occupancy of the interstitial site is about 3\% higher than the chemical Fe content obtained with ICP, $y \equiv y_{ICP} \approx y_{NPD} - 0.03$.

\section{Results and discussion}

%
\begin{figure}[!t]
\includegraphics[width=0.9\linewidth]{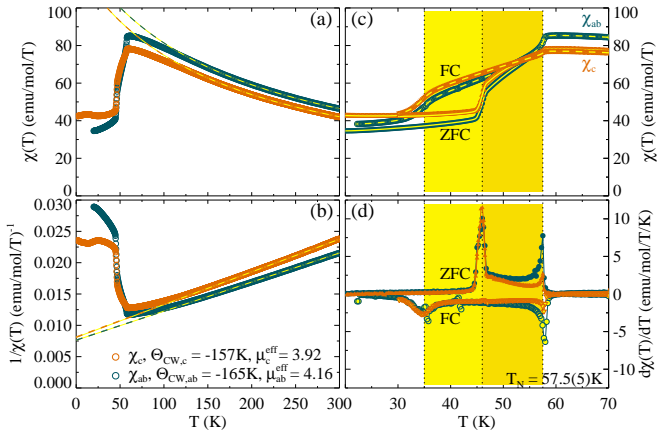}
\caption{(a) Temperature dependence of the ZFC static magnetic susceptibility of the Fe$_{1.09}$Te single crystal measured in DC magnetic field $B = 0.1$ T applied in the $ab-$plane, $\chi_{ab} (T)$, and along the $c-$axis, $\chi_{c} (T)$. (b) inverse susceptibility, illustrating the Curie-Weiss linear asymptotics at high temperatures. Dashed lines are Curie-Weiss fits.
(c), (d) The ZFC and the field-cooled (FC) susceptibilities and their derivatives. The shaded region shows phase with temperature-dependent magnetic incommensurability $\bQ_m$, the hysteretic range of the lock-in transition is light-shaded.}
\label{Fig1:Chi}
\vspace{-0.2in}
\end{figure}
%

%
\begin{figure}[!t]
\vspace{-0.25in}
\includegraphics[width=1.\linewidth]{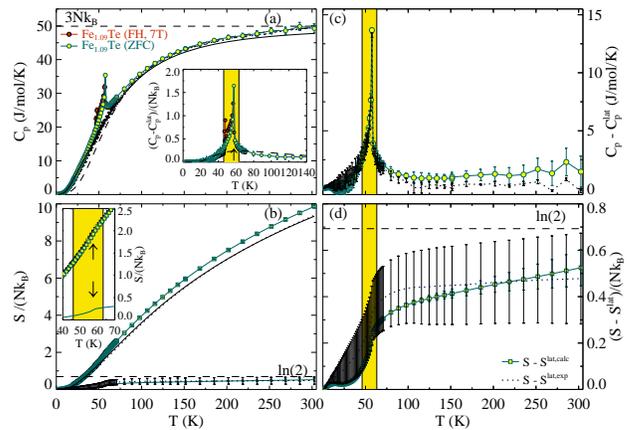}
\caption{(a) ZFC (light-filled) and FC (dark-filled symbols) specific heat capacity of Fe$_{1.09}$Te. The solid curve is the estimated lattice vibrational contribution, the dashed curve is the Debye fit in the $T > 130$ K range, the inset shows the corresponding net magnetic specific heat. (b) The total (light-filled symbols) and the net magnetic (connected points with error bars) entropy. Curves show the estimated lattice contribution. Inset expands the region near $T_N = 57.5(5)$ K, which is marked by the arrows. (c) and (d) show the net magnetic heat capacity and the magnetic entropy, respectively. Symbols and error bars connected by the dashed lines result from two different estimates of the lattice $C_p$ described in the text.
 }
\label{Fig2:Cp}
\vspace{-0.2in}
\end{figure}
%

Figure \ref{Fig1:Chi}(a) shows the temperature dependence of the magnetic susceptibility measured upon heating the zero-field-cooled (ZFC) sample. For $T \gtrsim 100$~K, it obeys the Curie-Weiss (CW) law,
\begin{equation}
 \chi_{\alpha}(T) = N_A \frac{(\mu^{\rm eff}_{\alpha})^2}{3k_B(T-\Theta_{{\rm CW},\alpha})},
\end{equation}
where $N_A$ is Avogadro's number, $k_B$ is Boltzmann's constant, $\alpha = ab, c$ indexes the field direction, and $\mu^{\rm eff}_{\alpha}$ and $\Theta_{{\rm CW},\alpha}$ are the effective paramagnetic moment and the Curie-Weiss temperature. This behavior is consistent with previous studies \cite{Li2009,Chen2009,HuPetrovic2009} and is best revealed by plotting $1/\chi_{ab,c}$ as in Fig.~1(b). CW fits to our data in the range $\Theta_{\rm CW} \lesssim T \lesssim  300$~K yield large effective magnetic moments, $\mu^{\rm eff}_\alpha = g_\alpha \mu_B \sqrt{S(S+1)}$, which are consistent with Fe atoms having local spins $S = 3/2$ with slightly anisotropic Lande $g-$factors, $g_{ab} \approx 2.15$ and $g_{c} \approx 2.02$. The CW temperatures are negative, corresponding to a dominant antiferromagnetic interaction, and are also very slightly anisotropic, with $\Theta_{{\rm CW},ab} = -165$~K and  $\Theta_{{\rm CW},c} = -157$~K. Remarkably, the magnitude of $k_B \Theta_{{\rm CW},\alpha}$ is much smaller than the bandwidth of magnetic excitations \cite{Lipscombe2011,Zaliznyak2011}, which indicates strong competition between ferro- and antiferromagetism. The low-$T$ susceptibility and its derivative in Figure \ref{Fig1:Chi}(c),(d) clearly reveal a continuous magnetic phase transition at $T_N = 57.5(5)$ K. $\chi(T)$ has a cusp at $T_N$, rather than a first-order-like discontinuity, which is further corroborated by the $\lambda$-like (step) singularity in $d\chi/dT$. It is followed by another magnetic transition, at a lower temperature, showing ZFC-FC hysteresis in the 35--45~K range.

The continuous, second-order nature of magnetic ordering at $T_N$ is further corroborated by the heat capacity in Fig.~\ref{Fig2:Cp}, which shows a $\lambda-$type singularity. The change in magnetic entropy associated with the long-range magnetic ordering at $T_N$ is very small, Fig. \ref{Fig2:Cp}(b),(d). It does not even reach $k_B \ln 2$ per Fe, which would correspond to freezing of a single Ising degree of freedom per Fe. This, together with a rather small ordered magnetic moment observed by neutron diffraction, $\langle \mu \rangle \lesssim 2\mu_B$,\cite{Bao2009,Rodriguez2011,Li2009,Martinelli2010,Zaliznyak2011} shows that long-range order (LRO) is weak. The most likely reason for the weakness of magnetic order is frustration arising from the competition of ferro- and antiferromagnetic, nearest- and further-neighbor interactions, which is also indicated by the small $\Theta_{\rm CW}$. The difference between the ZFC and the FC $C_p(T)$, as well as that measured upon heating the field-cooled sample in a magnetic field of $B = 7$~T, is very small, suggesting that the hysteretic transition observed in susceptibility at 35--45~K is some sort of spin realignment, involving negligible change in magnetic entropy.

\begin{figure}[!t]
\vspace{0.125in}
\includegraphics[width=1.2\linewidth]{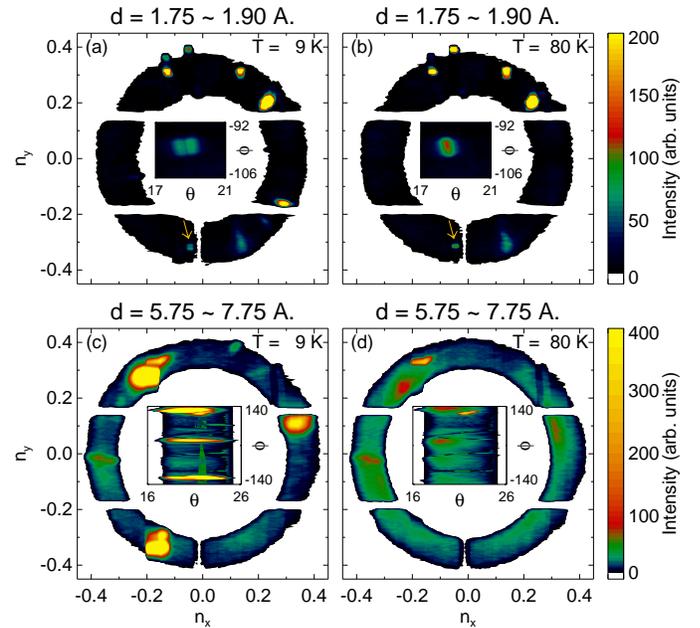}
\vspace{0.1in}
\caption{Contour maps of the scattered neutron intensity on the detector bank, as a function of the scattering direction, ${\bf n}$, at $T = 9$ K (a), (c) and $T = 80$ K (b), (d). $(n_x,n_y)$ are projections of the unit vector ${\bf n}$ pointing from sample to a detector pixel onto the plane perpendicular to the incident neutron beam direction, ${\bf k}_i$. Top panels, (a), (b), show the range of $d-$spacing dominated by structural scattering, and reveal the splitting of the (201) nuclear Bragg reflection on cooling. Bottom panels, (c), (d), show the large$-d$ range dominated by magnetic scattering, which transforms from a set of well-defined peaks at 9 K, to a ring-like feature typical of a liquid at 80 K. Insets show parts of the same data (marked by arrow) in angular coordinates, $(\theta,\phi)$, where $n_x = \sin\theta \cos\phi$, $n_y = \sin\theta \sin\phi$. The intensities have been window-averaged over the window of $3 \times 3$ detector pixels.
 }
\label{Fig3a:Laue}
\end{figure}

Figure~\ref{Fig3a:Laue} presents an overview of our neutron diffraction data at two temperatures, $T = 9$K and 80 K, in the form of Laue patterns on the detector bank. Each detector pixel is parameterized by the unit vector ${\bf n}(\theta,\phi)$ specifying its direction from the sample position. Although rather un-intuitive alignment of the sample reciprocal space with respect to the incident beam direction ($\approx 24^\circ$ rotation around ${\bf k}_i$ plus $\approx 45^\circ$ rotation around the vertical axis) results in an un-obvious pattern of equivalent Bragg peaks, reflections can be easily identified by their $d-$spacings. Panels (a), (b) show intensities corresponding to elastic scattering for the range of $d-$spacings from 1.75 \AA\ to 1.9 \AA, which is dominated by the lattice nuclear scattering. As magnified in the insets, a single Bragg spot corresponding to the $(2,0,1)$ lattice reflection seen at 80 K is clearly split into two spots at $T = 9$ K.

Intensities in the bottom panels (c), (d) of Fig.~\ref{Fig3a:Laue}, corresponding to 5.75 \AA $\leq d \leq $ 7.75 \AA, are mainly magnetic scattering. At the base temperature of 9 K it appears in the form of well-defined peaks near $(\pm 1/2,0,\pm 1/2)$ positions. Their significant angular size on detector results from the combination of deteriorating experimental resolution at small wave vectors (large $d-$spacings) and, to some extent, from contribution of quasielastic and inelastic diffuse scattering, which is also collected in this measurement. At $T = 80$ K only a small intensity modulation highlighting former peak positions remains, while magnetic intensity appears distributed over a ring of scattering. Such pattern is characteristic of a liquid.

\begin{figure}[!t]
\includegraphics[width=0.9\linewidth]{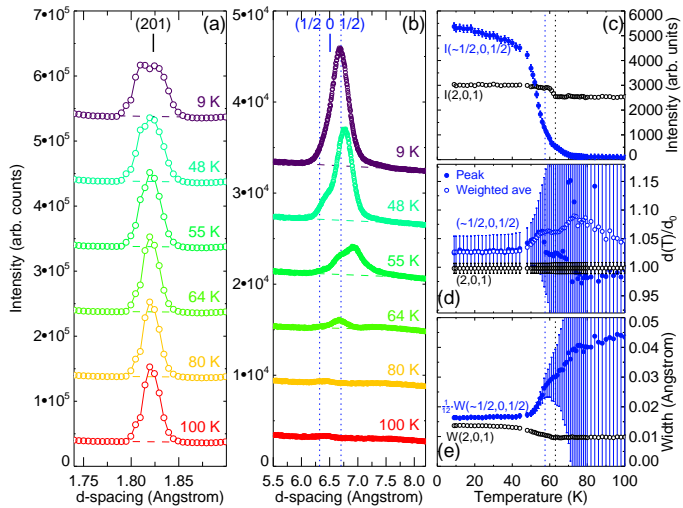}
\vspace{0.1in}
\caption{(a) Representative scans through the lattice Bragg peak $(2,0,1)$ and (b) the magnetic scattering near $\bQ = (0.5,0,0.5)$. The data at different T are vertically offset for presentation. (c) The integral intensity of the magnetic (filled blue symbols) and the lattice (open black symbols) scattering; (d) the maximum intensity position (filled blue symbols) and the intensity-weighted average position (open blue symbols) of magnetic peak and the intensity-weighted average of the lattice $(2,0,1)$ peak (open black symbols); (e) the width (intensity-weighted mean-square deviation of the measured points from the peak center) of magnetic (filled blue symbols) and the lattice (open black symbols) scattering.
 }
\label{Fig3:LaueTdep1}
\end{figure}

\begin{figure}[!t]
\includegraphics[width=0.9\linewidth]{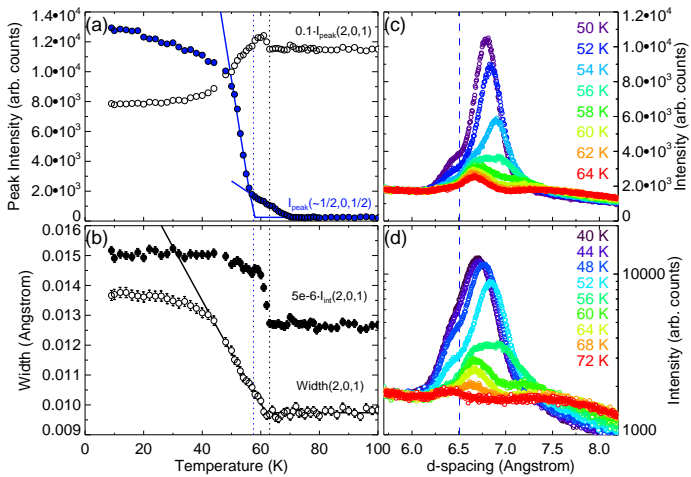}
\vspace{0.05in}
\caption{(a) The (maximum) peak intensity of the lattice (open) and the magnetic (filled symbols) scattering. (b) The integral intensity (filled symbols) and the width (intensity-weighted mean-square deviation of the measured points from the peak center, open symbols) of the $(2,0,1)$ lattice Bragg peak. Solid lines are fits to $\sim |T_{s,N}-T|^{2\beta}$ dependence. Dotted vertical lines show T$_s$ and T$_N$. (c) and (d) show representative magnetic scattering data at several temperatures on linear and logarithmic scale, respectively. Dashed vertical lines show $d-$spacing for $\bQ = (0.5,0,0.5)$.
 }
\label{Fig4:LaueTdep2}
\vspace{-0.25in}
\end{figure}

The competition between different magnetically ordered states, confirming the frustrated nature of magnetic interactions, is further revealed by the temperature dependencies of quasi-Laue neutron data in Figs.~\ref{Fig3:LaueTdep1} and \ref{Fig4:LaueTdep2}. They show intensities, obtained by integrating patterns, such as in Fig. \ref{Fig3a:Laue}, over the whole angular range of the detector array, as a function of $d-$spacing.  Magnetic peak profiles near the $\bQ = (1/2,0,1/2)$ position, Fig.~\ref{Fig3:LaueTdep1}(b) and Fig.~\ref{Fig4:LaueTdep2}(c,d), show complex structure evolving with temperature. Broad diffuse scattering around $d[\bQ \approx (0.4,0,1/2)] \approx 7.6$~\AA\  and a narrower peak near $d[\bQ \approx (0.51,0,1/2)] \approx 6.4$~\AA\ coexist for $T \gtrsim 70$ K. Below $\approx 70$ K they yield to a peak at $d[\bQ \approx (0.485,0,1/2)] \approx 6.65$ \AA, whose intensity grows roughly linearly with the decreasing $T$. Then, below $\approx 57.5$ K, a new dominant peak emerges at $d[\bQ \approx (0.46,0,1/2)] \approx 6.9$ \AA. Its position changes upon cooling, and finally saturates at  $d[\bQ \approx (0.48,0,1/2)] \approx 6.7 $ \AA\ below $\approx 45$ K. At the same time, a smaller peak appears at  $d[\bQ \approx (0.52,0,1/2)] \approx 6.3 $ \AA.

%
%

%
%

%
%

While very little, if any of these behaviors could be identified in the integrated magnetic intensity shown in Fig.~\ref{Fig3:LaueTdep1}(c), they are clearly observable in the temperature dependence of the peak maximum intensity, Fig.~\ref{Fig4:LaueTdep2}(a), and its position and width, Fig.~\ref{Fig3:LaueTdep1}(d), (e). In particular, the appearance of the new dominant magnetic component at $\approx 57.5$ K is most clear from an abrupt shift of the maximum intensity position in Fig.~\ref{Fig3:LaueTdep1}(d). Since it appears at the temperature where magnetic order is observed in susceptibility and heat capacity data, we identify this peak as a magnetic Bragg reflection associated with magnetic LRO. Fitting the peak intensity for $T \gtrsim 50$ K in Fig.~\ref{Fig4:LaueTdep2}(a) to an order-parameter-like dependence, $I(T) \sim (T_N-T)^{2\beta}$, we obtain $T_N = 57.5(5)$ K and $2\beta = 1.0(1)$, consistent with the mean-field, linear $I(T)$ behavior. The average peak position and its effective width in Fig.~\ref{Fig3:LaueTdep1}(d) and (e) are governed by the structured, multi-component nature of magnetic scattering and have very large error bars at high $T$, where the net magnetic intensity is small. Note that error bars on the maximum intensity position, which are shown by the closed symbols in Fig.~\ref{Fig3:LaueTdep1}(d), are much smaller, of the order of the symbol size.

Since fitting the overlapping peaks is sensitive to fitting ranges and constraints, peak parameters in Figs. \ref{Fig3:LaueTdep1} and \ref{Fig4:LaueTdep2} were evaluated directly from the measured intensities. The peak integral intensity, position, and width, were obtained by numerical integration of the measured intensity, the intensity-weighted position, and the mean-square deviation, respectively, upon subtracting the linear background interpolated between the edges of data ranges shown in Fig.~\ref{Fig3:LaueTdep1}(a) and (b). Parameters of the $(2,0,1)$ structural Bragg reflection, whose splitting at low temperature reveals the lattice distortion that reduces the symmetry from HTT, Fig.~\ref{Fig3:LaueTdep1}(a), were obtained in the same way. The position of magnetic and lattice peaks in Fig.~\ref{Fig3:LaueTdep1}(d) are given relative to the nominal positions, $d[\bQ = (1/2,0,1/2)]$ and $d[\bQ = (2,0,1)]$, respectively, in the HTT lattice with $a = 3.813$ \AA, $c = 6.24$ \AA.

While the evolution of the $(2,0,1)$ reflection between a single peak at 100 K and a two-peak structure at $\approx 9$ K seems gradual (mainly due to the experimental resolution), both the integral and maximum peak intensities in Fig.~\ref{Fig3:LaueTdep1}(c) and Fig.~\ref{Fig4:LaueTdep2}(a), as well as the peak width in Fig.~\ref{Fig4:LaueTdep2}(b), immediately reveal the structural phase transition at $\approx 63$ K. The width directly probes the order parameter -- the splitting of the $(2,0,1)$ peak. Fitting it to an order-parameter-like dependence in $T \gtrsim 48$ K range, we obtain $T_s = 63(1)$ K and $2\beta = 1.0(1)$, again consistent with the mean-field behavior. The intensity is an indirect probe, and its temperature dependence is governed by the combination of splitting and extinction, which is why the two dependencies in Fig.~\ref{Fig4:LaueTdep2}(b) differ. The lattice Bragg intensity measured on a large single crystal is reduced as a result of neutron beam extinction within the crystal. It is therefore very sensitive to $T_s$, where the crystal's mosaic structure changes due to the appearance of domains associated with the lowering of the HTT symmetry \cite{Marty2011}. While our present data do not allow us to distinguish between the monoclinic $P21/m$ and the orthorhombic $Pnmm$ structures,\cite{LynnDai2009} our supplementary powder diffraction measurements on a sample with similar $y$ indicate the $P21/m$ structure.

\section{Summary and conclusions}

To summarize, our analysis of neutron Bragg scattering establishes the structural phase transition, lowering the HTT lattice symmetry at $T_s = 63(1)$ K, as the first instability which occurs in Fe$_{1.1}$Te upon cooling. This transition is continuous and involves only very small structural changes: the $(2,0,1)$ peak splitting, $\delta d/d_0$, at 9 K is only $\sim 0.5$ \%. Moreover, this transition does not show up neither in heat capacity, Fig.~\ref{Fig2:Cp}, nor in bulk magnetic susceptibility, Fig.~\ref{Fig1:Chi}, indicating negligible entropy change and nearly complete decoupling between this structural change and the magnetic order. Close examination of the magnetic scattering in Figs.~\ref{Fig3:LaueTdep1} and \ref{Fig4:LaueTdep2} further corroborates this observation: its complex temperature evolution shows no visible anomaly at $T_s$.

Thus, the lattice distortion, which occurs in Fe$_{1.1}$Te at $T_s = 63(1)$~K, is not induced by the long-range antiferromagnetic ordering. Neither does it immediately lead to magnetic LRO, which only follows at a $\approx 10\%$ lower temperature, $T_N = 57.5(5)$~K. Magnetic order is \emph{weak}. It accounts for freezing of only $\lesssim 25\%$ of ($k_B \ln 4$/Fe) paramagnetic entropy of $S=3/2$ spins, implicated in the CW behavior of magnetic susceptibility at $T \gtrsim 100$ K. While the present quasi-Laue data is not suitable for the absolute normalization and refinement of the LRO moment involved in Bragg scattering, complementary monochromatic beam measurements \cite{Zaliznyak2011} indicate an ordered moment $\sim 1.4 \mu_B$ in our Fe$_{1.1}$Te crystal. This agrees with rather small values of the ordered magnetic moment, $\langle \mu \rangle \lesssim 2 \mu_B$, observed by neutron diffraction in Fe$_{1+y}$Te materials \cite{Bao2009,Rodriguez2011,Li2009,Martinelli2010}.

Magnetic LRO in our sample is not simple ``bicollinear'' type, but is an incommensurate structure, whose period varies with temperature. Upon cooling, it undergoes what looks like a lock-in transition with significant FC-ZFC hysteresis in magnetic susceptibility, as its propagation vector saturates at a low-$T$ value $\bQ \approx (0.48,0,1/2)$. In real space, such structure can be visualized by introducing a certain amount of ``tricollinear'' defects, whose density decreases with temperature. These defects can be viewed as randomly inserted lines of corner-sharing square plaquettes with ferromagnetically co-aligned spins. It was recently found that such plaquettes govern low-energy spin dynamics in Fe$_{1.1}$Te \cite{Zaliznyak2011}. Such proliferation of defects highlights magnetic frustration, where a number of different magnetic ground states, including the plaquette state, have nearly equal energy.\cite{Yin2010} This situation is further demonstrated by the temperature evolution of the magnetic neutron intensity, where, upon cooling, one observes competition of a number of states with different propagation vectors. We note that incommensurate magnetic orderings were also observed in a number of Se-doped compositions, Fe$_{1+y}$Te$_{1-x}$Se$_x$. \cite{Khasanov2009,Bendele2010,Babkevich2010}

Finally, our observations suggest that the weak lattice distortion and antiferromagnetic LRO are only symptomatic of the much stronger interactions driving the low-temperature physics in iron telluride. This leads us to question the applicability of traditional theoretical approaches, which treat electronic and magneto-structural properties in terms of expansions around the low-$T$ ordered states. The low-temperature physics in Fe$_{1+y}$Te, and, perhaps, in ferropnictides, too, is likely governed by rather high-energy degrees of freedom, such as the temperature-dependent orbital hybridization and interaction of local spins with itinerant electrons.\cite{Yin2010,Yin2011} Upon being properly integrated, these interactions should render an effective low-energy model governing the magneto-structural transition. Is there such a model, which could explain the extreme sensitivity to small amount of doping and the multi-critical phase diagrams common to Fe$_{1+y}$Te and 122 pnictides?

The effect of the small Fe off-stoichiometry, $y$, on the electronic structure of Fe$_{1+y}$Te being negligible \cite{Rodriguez2011}, Fe interstitials/vacancies can simply be viewed as a source of random magnetic and electric fields, which frustrate magnetic correlations, locally impact lattice distortion, and change the inter-layer structural and magnetic couplings. An effective description of the spontaneous HTT symmetry breaking in layered systems under these conditions is provided by the quasi two-dimensional (2D) anisotropic random field Ising model (ARFIM) considered by Zachar and Zaliznyak.\cite{ZacharZaliznyak2003} While this model was originally proposed for the superlattice formation associated with charge ordering in layered perovskites at half-doping, it is straightforwardly transplanted to the case of an orthorhombic distortion of the HTT lattice. The two possible choices of distortion at a given lattice site, which are related by the $90^\circ$ rotation, now play the role of an Ising variable. The effective ferromagnetic interaction accounts for the strain energy, arising where these two different states are adjacent to each other, and is strongly anisotropic (quasi-2D) in layered systems. The most important property of the ARFIM model is in its extreme sensitivity to very small amounts of disorder in the quasi-two-dimensional limit, which is inherited from the disordered nature of RFIM in 2D.\cite{ZacharZaliznyak2003} In fact, recent studies have found that ARFIM can account for signatures of nematicity observed in cuprates,\cite{LohCarlsonDahmen2010} and thus is probably also a good starting point for understanding similar findings in ferropnictides \cite{Nakajima2011}.

A natural generalization of the ARFIM\cite{ZacharZaliznyak2003} to cases where the monoclinic lattice distortion and/or strong magneto-elastic coupling result in four choices of the ground state, such as for the magneto-structural transitions in 122 ferropnictides and 11 chalcogenides, is provided by the four-state ($q=4$) anisotropic random field Potts model (ARFPM). Similarly to the ARFIM, the ordering in the ARFPM is governed by the 3D fixed point of the random field Potts model (RFPM). In RFPM, however, it is of the first order for small disorder.\cite{BlankensteinShapirAharony1984} A multicritical point, where the transition changes from first-order into a continuous one with the increasing random field strength, is expected in this case.\cite{BlankensteinShapirAharony1984} Hence, such a model could provide natural explanation for our current results, suggesting that such a multicritical point exists in Fe$_{1+y}$Te, as well as for the recent findings in 122 ferropnictides.\cite{Nakajima2011,ZhaoDai2008,LiDai2009,ChuFisher2009,Rotundu2010,Nandi2010,Ni2010,Kim2011,Marty2011}

\begin{acknowledgments}

We acknowledge discussions with and valuable assistance on different stages of this project from C. Petrovic, G. Xu, E. Carlson, and A. Tsvelik.
This work was supported by the Materials Sciences and Engineering Division, Office of Basic Energy Sciences (BES), US Department of Energy (DOE), under Contract No.\ DE-\-AC02-\-98CH10886. The work at the SNS was sponsored by the Scientific User Facilities Division, Office of BES, US DOE, under Contract No. DE-AC05-00OR22725.

\end{acknowledgments}


\end{document}